# Crowdsourcing open citations with CROCI
# An analysis of the current status of open citations, and a proposal

Ivan Heibi[1], Silvio Peroni[1] and David Shotton[2]

[1] *{ivan.heibi2, silvio.peroni}@unibo.it*
Digital Humanities Advanced Research Centre (DHARC), Department of Classical Philology and Italian Studies, University of Bologna, via Zamboni 32, 40126 Bologna (Italy)

[2] *david.shotton @oerc.ox.ac.uk*
Oxford e-Research Centre, University of Oxford, 7 Keble Rd, Oxford OX1 3QG (United Kingdom)

**Abstract**
In this paper, we analyse the current availability of open citations data in one particular dataset, namely COCI (the OpenCitations Index of Crossref open DOI-to-DOI citations; http://opencitations.net/index/coci) provided by OpenCitations. The results of these analyses show a persistent gap in the coverage of the currently available open citation data. In order to address this specific issue, we propose a strategy whereby the community (e.g. scholars and publishers) can directly involve themselves in crowdsourcing open citations, by uploading their citation data via the OpenCitations infrastructure into our new index, CROCI, the Crowdsourced Open Citations Index.

**Introduction**

The availability of open scholarly citations – i.e. citation data that are *structured*, *separate*, *open*, *identifiable* and *available* (Peroni and Shotton, 2018) – is a public good, which is of intrinsic value to the academic world as a whole (Shotton, 2013; Peroni et al. 2015; Shotton, 2018), and is particularly crucial for the scientometrics and informetrics community, since it supports reproducibility (Sugimoto et al., 2017) and enables fairness in research by removing such citation data from behind commercial paywalls(Schiermeier, 2017). Despite the positive early outcome of the Initiative for Open Citations (I4OC, https://i4oc.org/), namely that almost all major scholarly publishers now release their publication reference lists, with the result that more than 500 million citations are now open via the Crossref API (https://api.crossref.org), and despite the related ongoing efforts of sister infrastructures and initiatives such as OpenCitations (http://opencitations.net) and WikiCite/Wikidata (https://www.wikidata.org), many scholarly citations are not freely available. While these initiatives have the potential to disrupt the traditional landscape of citation availability, which for the past half-century has been dominated by commercial interests, the present incomplete coverage of open citation data is one of the most significant impediments to open scholarship (van Eck et al., 2018).

In this work, we analyse the current availability of open citations data within one particular dataset, namely COCI (the OpenCitations Index of Crossref open DOI-to-DOI citations; http://opencitations.net/index/coci). This dataset is provided by OpenCitations (Peroni et al., 2015), a scholarly infrastructure organization dedicated to open scholarship and the publication of open bibliographic and citation data by the use of Semantic Web (Linked Data) technologies. Launched in July 2018, COCI is the first of the Indexes proposed by OpenCitations (http://opencitations.net/index) in which citations are exposed as first-class data entities with accompanying properties. It has already seen widespread usage (over nine hundred thousands API calls since launch, with half of these in January 2019), and has been adopted by external services such as VOSviewer (van Eck and Waltman, 2010).

In particular, in this paper we address the following research questions (RQs):

1. What is the ratio between open citations vs. closed citations within each category of scholarly entities included in COCI (i.e. journals, books, proceedings, datasets, and others)?

2. Which are the top twenty publishers in terms of the number of open citations received by their own publications, according to the citation data available in COCI?
3. To what degree are the publishers highlighted in the previous analysis themselves contributing to the open citations movement, according to the data available in Crossref?

The results of these analyses show a persistent gap in the coverage of the currently available open citation data. To address this specific issue, we have developed a novel strategy whereby members of the community of scholars, authors, editors and publishers can directly involve themselves in crowdsourcing open citations, by uploading their citation data via the OpenCitations infrastructure into our new index, **CROCI, the Crowdsourced Open Citations Index**.

**Methods and material**

To answer the RQs mentioned above, we used open data and technologies coming from various parties. Specifically, the open CC0 citation data we used came from the CSV dump of most recent release of COCI dated 12 November 2018 (OpenCitations, 2018), which contains 449,840,503 DOI-to-DOI citation links between 46,534,705 distinct bibliographic entities. The Crossref dump we used for the production of this most recent version of COCI was dated 3 October 2018, and included all the Crossref citation data available at that time in both the 'open' dataset (accessible by all) and the 'limited' dataset (accessible only to users of the Crossref Cited-by service and to Metadata Plus members of Crossref, of which OpenCitations is one – for details, see https://www.crossref.org/reference-distribution/).

We additionally extracted information about the number of closed citations to each of the 99,444,883 DOI-identified entities available in the October 2018 Crossref dump. This number was calculated by subtracting the number of open citations to each entity available within COCI from the value "is-referenced-by-count" available in the Crossref metadata for that particular cited entity, which reports all the DOI-to-DOI citation links that point to the cited entity from within the whole Crossref database (including those present in the Crossref 'closed' dataset).

Furthermore, we extracted the particular publication type of each entity, so as to identify it either as a journal article, or as a book chapter, etc. We determined these publication types for all the DOI-identified entities available in the Crossref dump we used. We then identified the publisher of each entity, by querying the Crossref API using the entity's DOI prefix. This allowed us to group the number of open citations and closed citations to the articles published by that particular publisher, and to determine the top twenty publishers in terms of the number of open citations that their own publications had received.

Finally, we again queried the Crossref API, this time using the DOI prefixes of the *citing* entities, to check the participation of these top twenty publishers in terms of the number of open citations they were themselves publishing in response to the open citation movement sponsored by I4OC. Details of all these analyses are available online in CC0 (Heibi et al., 2019).

**Results**

First (RQ1) we determined the numbers of open citations and closed citations received by the entities in the Crossref dump. All the entity types retrieved from Crossref were aligned to one of following five categories: journal, book, proceedings, dataset, other – the mapping between Crossref types and the five types we used in our analysis is illustrated in the description of the table "croci_types.csv" in (Heibi et al., 2019). The outcomes are summarised in Figure 1, where it is evident that the number of open citations available in COCI is always greater than the number of closed citations to these entities within the Crossref database to which COCI does not have access, for each of the publication categories considered, with the categories *proceedings* and *dataset* having the largest ratios.

Analysis of the Crossref data show that there are in total ~4.1 million DOIs that have received no open citations and at least one closed citation. Conversely, there are ~10.7 million DOIs that have received no closed citations and at least one open citation in COCI. Most of the papers in both these categories have received very few citations.

The outcome of the second analysis (RQ2) shows which publishers are receiving the most open citations. To this end, we considered all the open citations recorded in COCI, and compared them with the number of closed citations to these same entities recorded in Crossref. Figure 2 shows the top twenty publishers that received the greatest number of open citations. Elsevier is the first publisher according to this ranking, but it also records the highest number of closed citations received (~97M vs. ~105.5M). The highest ratio in terms of open citations vs. closed citations was recorded by IEEE publications (ratio 6.25 to 1), while the lowest ratio was for the American Chemical Society (ratio 0.73 to 1).

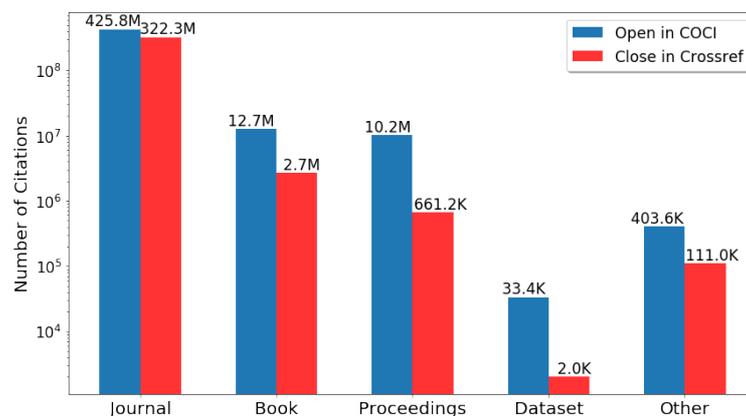

**Figure 1. The number of open citations (available in COCI) vs. closed citations (according to Crossref data) received by the cited entities within COCI, analyzed and grouped according to five distinct categories. [Note that the vertical axis has a logarithmic scale].**

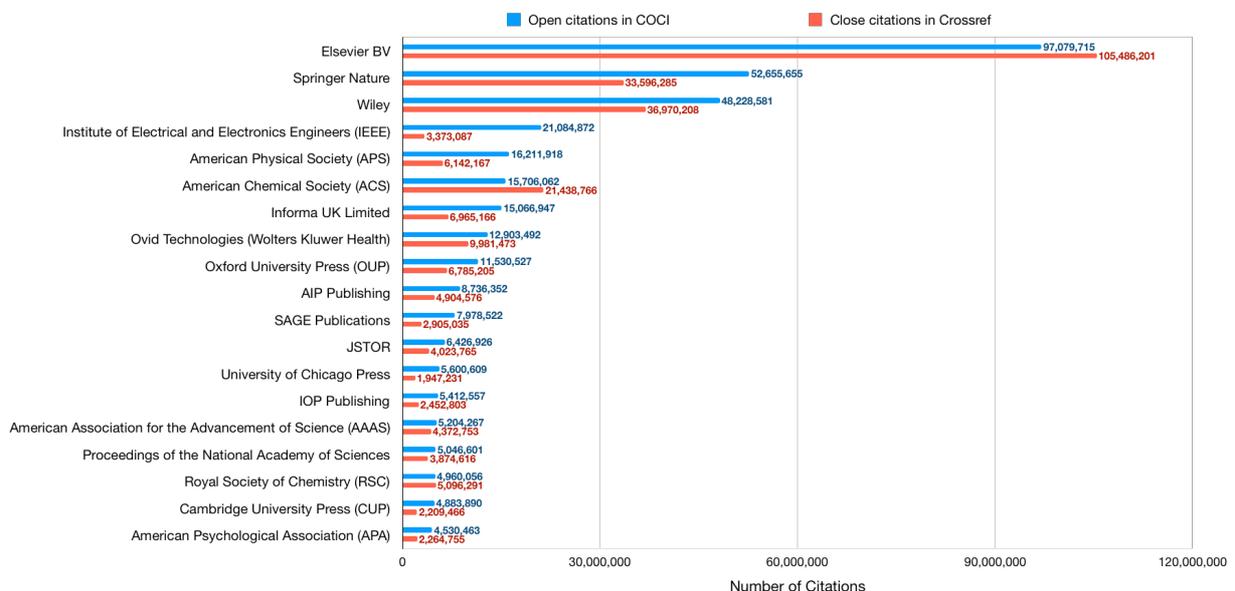

**Figure 2. The top twenty publishers sorted in decreasing order according to the number of open citations the entities they published have received, according to the open citation data within COCI. We accompany this count with the number of closed citations to the entities published by each of them according to the values available in Crossref.**

Considering the twenty publishers listed in Figure 2, we wanted additionally to know their current support for the open citation movement (RQ3). The results of this analysis (made by querying the Crossref API on 24 January 2019) are shown in Figure 3. Among the top ten publishers shown in Figure 2, i.e. those who themselves received the largest numbers of open citations, only five, namely Springer Nature, Wiley, the American Physical Society, Informa UK Limited, and Oxford University Press, are participating actively in the open publication of their own citations through Crossref.

It is noteworthy that JSTOR contributes very few references to Crossref, while the many citations directed towards its own holdings place JSTOR twelfth in the list of publishers receiving open citations (Figure 2). However, as the last column of Figure 3 shows, *all* the major publishers listed here are failing to submit reference lists to Crossref for a large number of the publications for which they submit metadata, that number being the difference between the value in the last column for that publisher and the combined values in the preceding three columns. JSTOR is the worst in this regard, submitting references with only 0.53% of its deposits to Crossref, while the American Physical Society is the best, submitting references with 96.54% of its publications recorded in Crossref.

Additional information about these analyses, including the code and the data we have used to compute all the figures, is available as a Jupyter notebook at https://github.com/sosgang/pushing-open-citations-issi2019/blob/master/script/croci_nb.ipynb.

| Publisher submitting references to Crossref | Closed | Limited | Open | Overall publications deposited | Total with references |
|---|---|---|---|---|---|
| Elsevier BV | 11,020,314 (65.70%) | 0 (0.00%) | 0 (0.00%) | 16,773,716 | 11,020,314 (65.70%) |
| Institute of Electrical and Electronics Engineers (IEEE) | 3,331,913 (79.06%) | 15,189 (0.36%) | 0 (0.00%) | 4,214,422 | 3,347,102 (79.42%) |
| American Chemical Society (ACS) | 496,855 (31.78%) | 0 (0.00%) | 0 (0.00%) | 1,563,601 | 496,855 (31.78%) |
| University of Chicago Press | 41,566 (9.02%) | 0 (0.00%) | 0 (0.00%) | 461,070 | 41,566 (9.02%) |
| Ovid Technologies (Wolters Kluwer Health) | 0 (0.00%) | 820,456 (40.20%) | 0 (0.00%) | 2,041,106 | 820,456 (40.20%) |
| IOP Publishing | 0 (0.00%) | 632,543 (76.25%) | 0 (0.00%) | 829,525 | 632,543 (76.25%) |
| American Psychological Association (APA) | 0 (0.00%) | 19,535 (2.73%) | 0 (0.00%) | 716,697 | 19,535 (2.73%) |
| Informa UK Limited | 0 (0.00%) | 15,632 (0.31%) | 3,021,271 (60.85%) | 4,965,446 | 3,036,903 (61.16%) |
| Springer Nature | 0 (0.00%) | 10,248 (0.08%) | 5,854,527 (45.12%) | 12,976,225 | 5,864,775 (45.20%) |
| Cambridge University Press (CUP) | 0 (0.00%) | 8,249 (0.40%) | 555,170 (26.59%) | 2,087,518 | 563,419 (26.99%) |
| SAGE Publications | 0 (0.00%) | 4,826 (0.19%) | 1,196,568 (47.14%) | 2,538,472 | 1,201,394 (47.33%) |
| Wiley | 0 (0.00%) | 0 (0.00%) | 5,698,571 (64.22%) | 8,874,184 | 5,698,571 (64.22%) |
| American Physical Society (APS) | 0 (0.00%) | 0 (0.00%) | 621,989 (96.54%) | 644,288 | 621,989 (96.54%) |
| Oxford University Press (OUP) | 0 (0.00%) | 0 (0.00%) | 583,329 (15.73%) | 3,707,847 | 583,329 (15.73%) |
| AIP Publishing | 0 (0.00%) | 0 (0.00%) | 562,840 (73.02%) | 770,812 | 562,840 (73.02%) |
| Royal Society of Chemistry (RSC) | 0 (0.00%) | 0 (0.00%) | 331,526 (52.58%) | 630,524 | 331,526 (52.58%) |
| Proceedings of the National Academy of Sciences | 0 (0.00%) | 0 (0.00%) | 77,621 (55.37%) | 140,176 | 77,621 (55.37%) |
| American Association for the Advancement of Science (AAAS) | 0 (0.00%) | 0 (0.00%) | 27,002 (9.43%) | 286,420 | 27,002 (9.43%) |
| JSTOR | 0 (0.00%) | 0 (0.00%) | 11,097 (0.53%) | 2,088,803 | 11,097 (0.53%) |

Figure 3. The contributions to open citations made by the twenty publishers listed in Figure 2, as of 24 January 2018, according to the data available through the Crossref API. The counts listed in the first three results columns of this table refers to the number of publications for which each publisher has submitted metadata to Crossref that include the publication's reference list, the categories *closed*, *limited* and *open* referring to publications for which the reference lists are not visible to anyone outside the Crossref Cited-by membership, are visible only to them and to Crossref Metadata Plus members, or are visible to all, respectively. Additional information on this classification of Crossref reference lists is available at https://www.crossref.org/reference-distribution/. The fourth results column in the table shows the total number of publications for which the publisher has submitted metadata to Crossref, whether or not those metadata include the reference lists of those publications, and the fifth results column shows the total number of publications for which the publisher *has* submitted the reference list with the other metadata. The percentage values given in parentheses show the percentage of publications in each category whose metadata submitted to Crossref includes the reference lists, these percentages being obtained by dividing the values in each column by the total number of publications for which that publisher has submitted metadata to Crossref shown in the fourth results column.

It should be stressed that a very large number of potentially open citations are totally missing from the Crossref database, and consequently from COCI, for the simple reason that many publishers, particularly smaller ones with limited technical and financial resources, but also all the large ones shown in Figure 3 and most of the others, are simply not depositing with Crossref the reference lists for any or all of their publications.

**Discussion**

According to the data retrieved, the open DOI-to-DOI citations available in COCI exceed the number of closed DOI-to-DOI citations recorded in Crossref for every publication category, as shown in Figure 1. The journal category is the one receiving the most open citations overall, as expected considering the historical and present importance of journals in most areas of the scholarly ecosystem. However, the number of closed citations to journal articles within Crossref is also of great significance, since these 322 million closed citations represent 43% of the total.

It is important to note that about one third of these closed citations to journal articles (according to Figure 2) are references to entities published by Elsevier, and that references from within Elsevier's own publications constitute the largest proportion of these closed citations, since Elsevier is the largest publisher of journal articles. Thus, Elsevier's present refusal to open its article references is contributing significantly to the invisibility of Elsevier's own publications within the corpus of open citation data that is being increasingly used by the scholarly community for discovery, citation network visualization and bibliometric analysis.

It is also worth mentioning the discrepancy between the citations available in COCI, which comes from the data contained in the open and limited Crossref datasets as of 3 October 2018, and those available within those same Crossref datasets as of 24 January 2019. The most significant difference relates to IEEE. While the citations present in COCI include those from IEEE publications to other entities prior to November 2018 (since in October 2018 its article metadata with references were present within the Crossref *limited* dataset), in November 2019 this scholarly society decided to close the main part of its Crossref references, and thus from that moment they became unavailable to Crossref Metadata Plus members such as OpenCitations, as highlighted in Figure 3. Thus, IEEE citations from articles whose metadata was submitted to Crossref after the date of this switch to *closed* can no longer be automatically ingested into COCI.

To date, the majority of the citations present in Crossref that are not available in COCI comes from just three publishers: Elsevier, the American Chemical Society and University of Chicago Press (Figure 3). In fact, considering the average value of 18.6 DOI-to-DOI citation links for each citing entity – calculated by dividing the total number of citations in COCI by the number of citing entities in the same dataset – these three publishers are holding more than 214 million DOI-to-DOI citations that could potentially be opened. (The IEEE citation data which was in the Crossref 'limited' category as of October 2018 are actually included in COCI, although those from that organization's more recent publications will no longer be, as mentioned above). We think it is deeply regrettable and almost incomprehensible that any professional organization, learned society or university press, whose primary mission is to serve the interests of the practitioners, scholars and readers it represents, should choose *not* open all its publications' reference lists as a public good, whatever secondary added-value services it chooses to build on top of the citations that those reference lists contain.

**CROCI, the Crowdsourced Open Citations Index**

The results of the Initiative for Open Citations (I4OC) have been remarkable, since its efforts have led to the liberation of millions of citations in a relatively short time. However, many more citations, the lifeblood of the scholarly communication, are still not available to the general public, as mentioned in the previous section. Some researchers and journal editors, in particular, have recently started to interact with publishers that are not participating in I4OC, in attempts to convince them to release their citation data. Remarkable examples of these activities are the petition promoted by Egon Willighagen (https://tinyurl.com/acs-petition) addressed to the American Chemical Society, and the several unsuccessful requests made to Elsevier by the Editorial Board of the Journal of Informetrics, which eventually resulted in the resignation of the entire Editorial Board on 10 January 2019 in response to Elsevier's refusal to address their issues (http://www.issi-society.org/media/1380/resignation_final.pdf).

To provide a pragmatic alternative that would permit the harvesting of currently closed citations, so that they could then be made available to the public, we at OpenCitations have created a new OpenCitations Index: **CROCI, the Crowdsourced Open Citations Index**, into which individuals identified by ORCiD identifiers may deposit citation information that they have a legal right to submit, and within which these submitted citation data will be published under a CC0 public domain waiver to emphasize and ensure their openness for every kind of reuse without limitation. Since citations are statements of fact about relationships between publications (resembling statements of fact about marriages between individual persons), they are not subject to copyright, although their specific textual arrangements within the reference lists of particular publications may be. Thus, the citations from which the reference list of an author's publication has been composed may legally be submitted to CROCI, although the formatted reference list cannot be. Similarly, citations extracted from within an individual's electronic reference management system and presented in the requested format may be legally submitted to CROCI, irrespective of the original sources of these citations.

To populate CROCI, we ask researchers, authors, editors and publishers to provide us with their citation data organised in a simple four-column CSV file ("citing_id", "citing_publication_date", "cited_id", "cited_publication_date"), where each row depicts a citation from the citing entity ("citing_id", giving the DOI of the cited entity) published on a certain date ("citing_publication_date", with the date value expressed in ISO format "yyyy-mm-dd"), to the cited entity ("cited_id", giving the DOI of the cited entity) published on a certain date ("cited_publication_date", again with the date value expressed in ISO format "yyyy-mm-dd"). The submitted dataset may contain an individual citation, groups of citations (for example those derived from the reference lists of one or more publications on a particular topic), or entire citation collections. Should any of the submitted citations be already present within CROCI, **these duplicates will be automatically detected and ignored**.

The date information given for each citation should be as complete as possible, and minimally should be the publication years of the citing and cited entities. However, if such date information is unavailable, we will try to retrieve it automatically using OpenCitations technologies already available. DOIs may be expressed in any of a variety of valid alternative formats, e.g. "https://doi.org/10.1038/502295a", "http://dx.doi.org/10.1038/502295a", "doi: 10.1038/502295a", "doi:10.1038/502295a", or simply "10.1038/502295a".

An example of such a CVS citations file can be found at https://github.com/opencitations/croci/blob/master/example.csv. As an alternative to submissions in CSV format, contributors can submit the same citation data using the Scholix format (Burton et al., 2017) – an example of such format can be found at https://github.com/opencitations/croci/blob/master/example.scholix.

Submission of such a citation dataset in CSV or Scholix format should be made as a file upload either to Figshare (https://figshare.com) or to Zenodo (https://zenodo.org). For provenance purposes, the ORCID personal identifier of the submitter of these citation data should be explicitly provided in the metadata or in the description of the Figshare/Zenodo object. Once such a citation data file upload has been made, the submitter should inform OpenCitations of this fact by adding an new issue to the GitHub issue tracker of the CROCI repository at https://github.com/opencitations/croci/issues.

OpenCitations will then process each submitted citation dataset and ingest the new citation information into CROCI. These CROCI citations will be made available at http://opencitations.net/index/croci using a REST API and a SPARQL endpoint, and will additionally be published periodically as data dumps in Figshare, all releases being under CC0 waivers. We propose in future to enable combined searches over all the OpenCitations indexes, including COCI and CROCI.

We are confident that the community will respond positively to this proposal of a simple method by which the number of open citations available to the academic community can be increased, in particular since the data files to be uploaded have a very simple structure and thus should be easy to prepare. In particular, we hope for submissions of citations from within the reference lists of authors' green OA versions of papers published by Elsevier, IEEE, ACS and UCP, and from publishers not already submitting publication metadata to Crossref, so as to address existing gaps in open citations availability. We look forward to your active engagement in this initiative to further increase the availability of open scholarly citations.